\documentclass[]{WileyASNA-v1}

\articletype{Proceedings}%

\received{August 2024}
\revised{October 2024}
\accepted{November 2024}

\usepackage{hyperref}
\hypersetup{
    colorlinks,
    breaklinks=true,
    linkcolor={red},
    citecolor={blue},
    urlcolor={blue}
    }
    
\newcommand{\E}[1]{\times 10^{#1}}  

\def\sun{{\odot}}   
\def\NS{\text{NS}}  

\def\phot{\text{ph}}	
\def\eff{\text{eff}}    

\def\Teff{{T}_\eff}         
\def\LRphot{L_\text{R,ph}}  
\def\Rns{{R}_\NS}           
\def\Mns{{M}_\NS}            

\def\yr{\text{yr}}      
\def\Msun{\text{M}_{\sun}}  
\def\Lsun{\text{L}_{\sun}}  
\def\Msunyr{\text{M}_{\sun}~\yr^{-1}}    

\def\km{\text{km}}

\def\K{\text{K}}

\defcitealias{HSJ2020}{HSJ2020}
\defcitealias{HSJ2023}{HSJ2023}
\defcitealias{Jose2010}{JMPI2010}

\usepackage[]{apacite}
\bibstyle{Wiley-ASNA.bst}

\begin{document}

\title{Mass-loss, composition and observational signatures of stellar winds from X-ray bursts}

\author[1,3]{Herrera, Yago}

\author[2,3]{Mu\~noz Vela, Daniel}

\author[2,3]{Sala, Gloria}

\author[2,3]{Jos\'e, Jordi}

\author[2,3]{Cavecchi, Yuri}

\authormark{Herrera \textsc{et al.}}

\address[1]{\orgname{Institute of Space Sciences - CSIC}, \orgaddress{\city{Cerdanyola del Vall\'es}, \state{Barcelona}, \country{Spain}}}

\address[2]{\orgdiv{Departament de Física, EEBE}, \orgname{Universitat Polit\'ecnica de Catalunya}, \orgaddress{\city{Barcelona}, \country{Spain}}}

\address[3]{\orgname{Institut d'Estudis Espacials de Catalunya},  \orgaddress{\city{Castelldefels},  \country{Spain}}}

\corres{Herrera, Yago \email{herrera@ice.csic.es}}


\abstract{ X-Ray bursts (XRBs) are powerful thermonuclear events on the surface of accreting neutron stars (NSs), which can synthesize intermediate-mass elements. 
Although the high surface gravity prevents an explosive ejection, a small fraction of the envelope may be ejected by radiation-driven winds. 
In our previous works, we have developed a non-relativistic radiative wind model and coupled it to an XRB hydrodynamic simulation.
We now apply this technique to another model featuring consecutive bursts.
We determine the mass-loss and chemical composition of the wind ejecta.
Results show that, for a representative XRB, about $0.1\%$ of the envelope mass is ejected per burst, at an average rate of $3.9 \E{-12}\,\Msun \texttt{yr}^{-1}$. Between $66\%$ and $76\%$ of the ejecta composition is $^{60}$Ni, $^{64}$Zn, $^{68}$Ge, $^{4}$He and $^{58}$Ni. We also report on the evolution of observational quantities during the wind phase and simulate NICER observations that resemble those of 4U 1820-40. 
}

\keywords{{stars: neutron}, {X-rays: bursts},  {stars: winds, outflows}, nucleosynthesis, abundances, {methods: numerical}}



\maketitle


\section{Introduction}\label{sect: intro} 

Type I X-ray bursts (XRBs) are powerful, recurrent thermonuclear explosions on the surface of accreting neutron stars in binary systems \citep{StrohmBild2003, KeekZand2008, GalloMunHartChak2008,JoseBook2016}. 
They occur when accreted material builds up under moderately degenerate conditions, triggering nuclear reactions and a thermonuclear runaway. 
This leads to a sharp increase in luminosity and the production of intermediate-mass elements, mainly around A=64   \citep{Woosley_2004,Fisker_2008,Jose2010}. 
A typical neutron star (NS), with an escape velocity near $\tfrac{2}{3}c$, allows only a limited envelope expansion before the nuclear reactions stop due to nuclear fuel consumption. 
However, at certain accretion rates, significant photospheric radius expansion (PRE) can occur, with luminosity reaching or exceeding the Eddington limit, and potentially leading to material ejection via a radiation-driven wind \cite[see][and references therein --\citetalias{HSJ2020}, from now on]{HSJ2020}. 
The contribution of XRBs to Galactic abundances is still debated and is one of the goals of this study. 
In addition, the study of XRB winds in PRE bursts can lead to the determination of neutron stars radii and masses, therefore providing constraints on the equation of state of NS matter \citep{Lattimer+Prakash-2006, Damen++1990,Steiner2010,Guver++2012,Sala++2012}, but several simple assumptions are usually taken for the determination of the NS mass (maximum luminosity at the Eddington limit) and radius (touch-down of photospheric radius at NS surface). 
The simulation of the NS envelope during XRB winds can help to reduce uncertainties and test the validity of those assumptions.
  
The last years have witnessed a renewed interest in the modeling of XRB winds. 
\cite{YuWeinberg2018} performed MESA \citep{MESA1} simulations of the hydrodynamics of the wind after the Eddington limit is reached; and \cite{Guichandut2021} studied the transition from static expanded envelopes to radiatively driven stellar wind and discussed the applicability of steady-state models. 

In \citetalias{HSJ2020} we determined the conditions for the presence of a radiative wind in a generic NS scenario, we explored the wind model parameter space in terms of energy and mass outflows ($\dot E, \dot M$), and analyzed possible predictions related to observable variables. 
Subsequently, in \cite{HSJ2023} --\citetalias[][from now on]{HSJ2023}--, we applied the wind model to a more realistic XRB scenario by coupling it to the results of a high resolution (200 shells) hydrodynamic simulation by \cite{Jose2010} --\citetalias{Jose2010} from now on--, which modeled the thermonuclear reactions and complex hydrodynamics of the XRB in the accreted envelope.
This way we could self-consistently simulate the evolution of the entire NS envelope, from the nuclear burning shells to the photosphere of the expanding wind, allowing us to study the evolution of observable quantities for the XRB-wind model, to obtain the composition of the envelope layers blown away by the wind, and ultimately to assess the possible contribution of XRBs to Galactic abundances. 

 As a natural continuation to the work presented in \citetalias{HSJ2023}, here we present the wind model coupled to another simulation of \citetalias{Jose2010}.
 While this time we use a lower resolution simulation (60 shells) than the one studied in \citetalias{HSJ2023}, it simulates four consecutive bursts in a continuous time sequence, accounting for the accretion between bursts and thus the natural chemical evolution. 
 Additionally, we were able to reconstruct the continuous evolution of the photospheric properties of each burst, featuring both the radiative wind and the wind-free phases, for which we have also simulated observations with the \textit{Neutron Star Interior Composition Explorer Mission} (\textit{NICER}).

\section{Models and methods}\label{sec2}
\label{sect: XRB_Wind}
 
  We provide here a summary of our models and methodology, and refer the reader to the original articles for details. 
  The radiative wind \citepalias[see][]{HSJ2020} is simulated by a stationary, non-relativistic model with spherical symmetry, assuming a fully-ionized perfect gas in local thermodynamic equilibrium (LTE). 
  The X-ray burst hydrodynamic models from \citetalias{Jose2010} couple a nuclear reaction network (324-isotopes linked by 1392 nuclear interactions), into a modified version of the \textit{SHIVA} hydrodynamic evolution code \cite[see][]{Jose-Hernanz-1998,JoseBook2016}. 
  The hydrodynamic code assumes spherical symmetry, Newtonian gravity, and energy transport by radiative diffusion and convection. 
  
  In order to match the two models, we search the grid of time-steps and radial-shells in the hydrodynamical models for points where the physical conditions to power a radiative stellar wind are fulfilled.
  Once suitable points are identified, a radiative wind model solution is calculated adjusting its free parameters to match the base conditions provided at the matching points by the hydrodynamic model. 
  We assume a radially uniform chemical abundance for the wind model, determined from the XRB hydrodynamic models at the matching point, and changing for each time-step.
  Most of the physical variables of the hydrodynamical model at the matching points between hydrodynamic and wind model show some degree of fluctuation and irregularity in time distribution, which are carried on to the wind model solutions. 
  A smoothing technique using local regression was applied, incorporating the matching error as a weight factor, so that the smoothing favors points with smaller error. 
  The matching technique between the XRB hydrodynamic model and the wind model is described in detail in \citetalias{HSJ2023}.
  
  Unlike the high-resolution single burst model presented in \citetalias{HSJ2023}, in the present work we focused on the four consecutive bursts simulated in the lower-resolution model of \citetalias{Jose2010} (60-shell model), for a neutron star with $\Mns =  1.4~\Msun$, $\Rns=13.1~\km$, accreting solar composition material at a rate of $ 1.75 \E{-9} \,\Msunyr$. Its main resulting properties are burst recurrence times ranging in about $[5-6.5] \text{ hr} $, burst duration of $[55-75] \text{ s} $, peak luminosity within $\sim [1-2] \E{5}~\Lsun$, and peak temperatures in $[1-1.25] \E{9}~\K$.

  Finally, once we obtained XRB-wind model predictions for observable quantities, we aim at providing a proof of the viability of the detection of the presence of winds by simulating observations with \textit{NICER}. 
  The full evolution of photospheric quantities, such as radiative luminosity ($\LRphot$), effective temperature ($\Teff$), photospheric radius ($r_\phot$), and element abundances have been considered throughout the entire duration of each burst, both for the wind and the wind-free phases. 
  While the photospheric values are well determined in the wind model (as they are the outer boundary conditions), some assumptions were needed for the XRB hydrodynamic models, since they focus mostly on nucleosynthesis in the inner layers.  
  An extrapolation technique was developed to obtain photospheric values of the XRB wind-free phases.
  We assumed the photosphere is located at $\tau^* =\kappa \rho r \simeq 8/3$, as prescribed by \cite{Kato1983}. 
  There, $\LRphot$ can be considered to be equal to the one in the last reported shell, since no further nuclear reactions are assumed to occur above it; the same can be said for element abundances, since convective and diffusive mixing are negligible.
  We assume the effective temperature at the photosphere obtained from Stephan-Boltzmann's law. 
 
\section{Results}
\label{sect: XRB-wind results}
  The search for the XRB-wind model matching points was successful for the XRB hydrodynamic simulation studied, i.e., for all bursts analyzed there was a brief phase of radiative wind and a consequent mass-loss. 
  For brevity, we show results for the second and third burst in the sequence (XRB-2 and XRB-3).
  The shell where the wind conditions were fulfilled in the hydrodynamic model (the matching point) followed a well defined progression: starting in the outer (cooler and less dense) layers, progressing to the inner (hotter and denser) ones as the wind further expands the photosphere, and returning to the outer ones again as the wind recedes.
  That is, we now have the full evolution of the wind phase, featuring both expansion and recession, in contrast to our higher resolution model in \citetalias{HSJ2023}, where we only had the receding part of the wind. 
  The resulting mass outflow, $\dot M$, consequently showed higher values at the inner matching layers, diminishing when the matching point lies at shallower depths. 

\subsection{Wind mass-loss and composition}
\label{sect: XRB-wind results mass-loss}

We integrated the mass outflow rate of each isotope, $\dot m_i(t)$, over the duration of the wind to calculate the total mass-loss for each species, $\Delta m_i$. 
The total mass ejected was $\Delta m ^\text{(XRB-2)} = 7.6\E{18}\ \texttt{g}$ , and $\Delta m ^\text{(XRB-3)}= 2.2\E{18}\ \texttt{g}$, a fraction of the order of $10^{-3}$ of the mass of the NS envelope. 
Given that the model has an average recurrence time of $5.6\ \texttt{hr}$, and assuming continuous activity, this results in a mean yearly output of $3.9 \E{-12}\,\Msun \texttt{yr}^{-1}$, which is $0.3\%$ of the mass-accretion rate.
These results are almost an order of magnitude smaller than the ones we found in \citetalias{HSJ2023}, mostly due to the shallower layers at which the wind conditions are met, in the XRB model analyzed here.

\begin{figure}
  \centering 
  \includegraphics[keepaspectratio=true,width=.9\columnwidth,clip=true,trim=0pt  75pt 0pt 40pt]{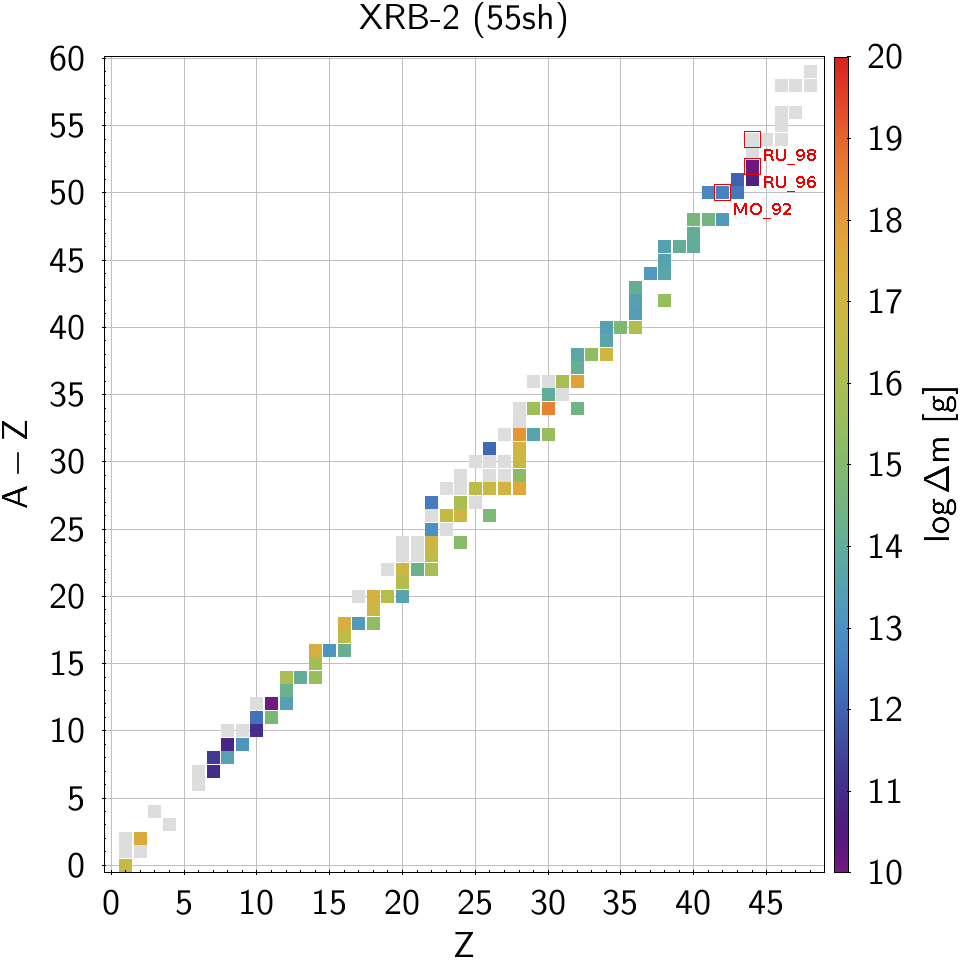} \\[2pt]
  \includegraphics[keepaspectratio=true,width=.9\columnwidth,clip=true,trim=0pt  0pt 0pt 40pt]{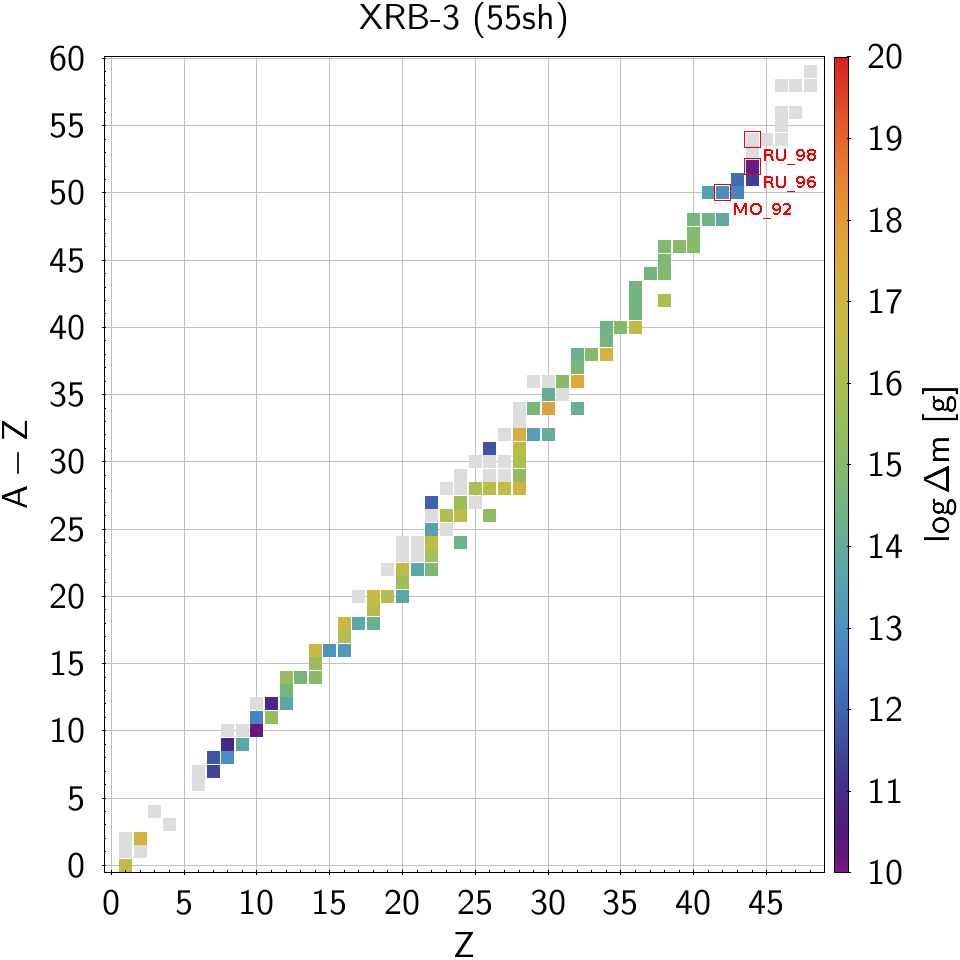}
  \caption{
    Total mass yield of stable isotopes from stellar wind in XRB-2 (top) and XRB-3 (bottom). 
  }
  \label{fig: XRB-Wind isotopes A-Z stable} 
\end{figure}

Figure \ref{fig: XRB-Wind isotopes A-Z stable} shows the total mass yield of stable isotopes produced in XRB-2 and 3 and ejected by the stellar wind, with yields color-coded for isotopes with $\Delta m_i > 10^{10}\ \texttt{g}$. 
The top ten isotopes are listed in Table \ref{tab: top isotopes}. 
Nearly $76\%$ of the total ejected mass in XRB-2 ($66\%$ in XRB-3) consists of the first five isotopes:  $^{64}\text{Zn}$, $^{60}\text{Ni}$, $^{68}\text{Ge}$, $^{56}\text{Ni}$, and $^{4}\text{He}$. These, along with the other isotopes in the table, account for over $88\%$ of the total mass ejected in XRB-2 ($83\%$ in XRB-3). 

The mass fractions of some isotopes vary noticeable between bursts. 
This can partly be attributed to the difference in available nuclear fuel in each consecutive burst, but also to the depth at which the XRB conditions match a wind solution (i.e., which layers are blown away by the wind).

\begin{table}
  \begin{center}
  \caption{
    Top ten stable isotopes after radioactive decay by mass yield, from stellar wind in XRB-2 and XRB-3.
  }
  \label{tab: top isotopes}
   \scriptsize
  XRB-2 \\[1ex]
  
  \begin{tabular}{lrrrr}
    \hline
	Sym     &   Z  &   A  &   Mass (g)                  & Mass fraction            \\
    \hline
    Zn	& $30$	& $64$	& $3.07\E{18}$	& $4.07\E{-01}$ \\
    Ni	& $28$	& $60$	& $1.35\E{18}$	& $1.80\E{-01}$ \\
    Ge	& $32$	& $68$	& $6.30\E{17}$	& $8.36\E{-02}$ \\
    Ni	& $28$	& $56$	& $3.41\E{17}$	& $4.52\E{-02}$ \\
    He	& $2 $	& $4 $	& $3.37\E{17}$	& $4.47\E{-02}$ \\
    S	& $16$	& $34$	& $2.42\E{17}$	& $3.21\E{-02}$ \\
    Si	& $14$	& $30$	& $1.96\E{17}$	& $2.60\E{-02}$ \\
    Ar	& $18$	& $38$	& $1.77\E{17}$	& $2.35\E{-02}$ \\
    Ti	& $22$	& $46$	& $1.43\E{17}$	& $1.89\E{-02}$ \\

    \hline
  \end{tabular}

   \vspace{2ex}

    XRB-3 \\[1ex]
  
  \begin{tabular}{lrrrr}
    \hline
	Sym     &   Z  &   A  &   Mass (g)                  &   Mass fraction           \\
    \hline
    Zn 	& $30$ 	& $64$ 	& $5.15\E{17}$ 	& $2.36\E{-01}$ \\
    Ge 	& $32$ 	& $68$ 	& $3.42\E{17}$ 	& $1.57\E{-01}$ \\
    Ni 	& $28$ 	& $60$ 	& $2.85\E{17}$ 	& $1.30\E{-01}$ \\
    He 	& $2$ 	& $4$ 	& $1.75\E{17}$ 	& $7.99\E{-02}$ \\
    Ni 	& $28$ 	& $56$ 	& $1.37\E{17}$ 	& $6.29\E{-02}$ \\
    Se 	& $34$ 	& $72$ 	& $1.34\E{17}$ 	& $6.11\E{-02}$ \\
    S 	& $16$ 	& $34$ 	& $7.62\E{16}$ 	& $3.48\E{-02}$ \\
    Si 	& $14$ 	& $30$ 	& $6.21\E{16}$ 	& $2.84\E{-02}$ \\
    Ar 	& $18$ 	& $38$ 	& $5.35\E{16}$ 	& $2.45\E{-02}$ \\
    \hline
  \end{tabular}
  \end{center}
\end{table}

\subsection{Observables and their correlations}
\label{sect: XRB-wind results observables}
  
 Another key aspect we aim to determine from the XRB-wind matching results is the evolution of observable features and their relationship to other predicted physical variables. 
 The study of the one-burst simulation with higher spatial resolution in \citetalias{HSJ2023} (the 200-shell model) does not capture the full wind evolution due to mismatches between the XRB hydrodynamic and the wind model assumptions. However, the analysis of the four consecutive bursts from the lower-resolution simulations (60-shell model) in the present work supplements that study and contributes to a more complete understanding.

\begin{figure*}

  \centering 
  \includegraphics[width=.8\linewidth]{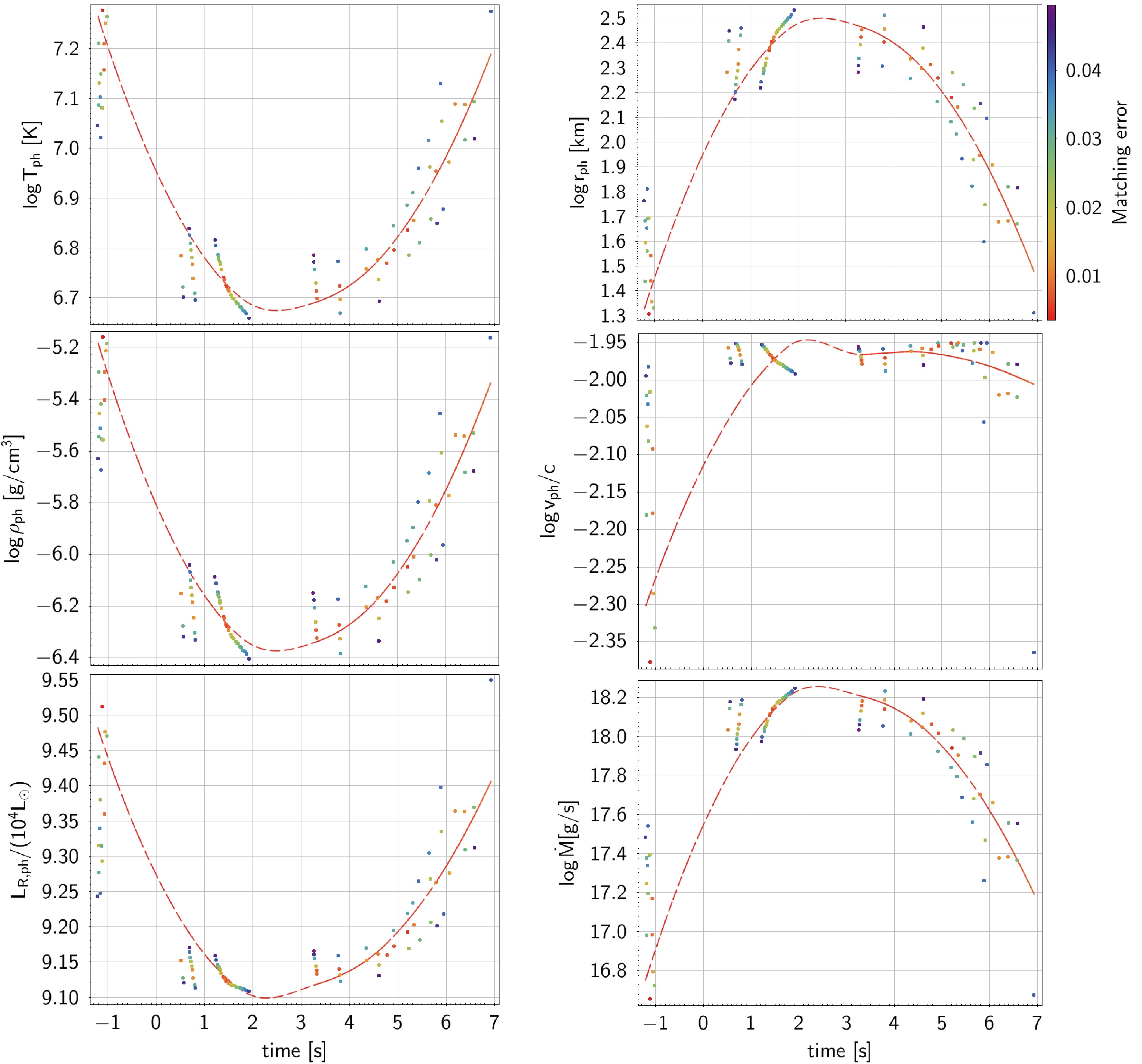}
  
  \caption{ 
    Time evolution of photospheric quantities (temperature, radius, density, wind velocity, and radiative luminosity, in reading order) and mass outflow during the wind phase, in XRB-2.
   Dots indicate values corresponding to wind model solutions that match XRB hydrodynamic model conditions, with a matching error indicated by the color scale. Lines show the predicted values, using a smoothing-interpolating technique. See \citetalias{HSJ2023} for further details.
    }
    \label{fig: XRB-Wind photosphere time smoothing} 
\end{figure*}

  The time evolution of photospheric quantities is shown in Fig. \ref{fig: XRB-Wind photosphere time smoothing}. 
  The photospheric radius expands from a few km above the NS core ($\Rns=13.1\ \texttt{km}$) to about $300 \ \texttt{km}$ and then recedes symmetrically towards the original expansion.
  Inversely, the photospheric temperature drops from about $2\E{7}\ \texttt{K}$ at the wind onset, cooling down to about $4.5 \E{6} \ \texttt{K}$ during the maximum wind expansion, and rising again afterwards. 
  The wind velocity remains around $\sim 0.01\ \texttt{c}$ after quickly rising at the wind onset.
  The photospheric luminosity decreases during the wind phase, in an opposite way to mass outflow $\dot M$. 
  The change is small, however, $\sim 5-6\%$, and negligible compared to the burst's luminosity rise.
  The average wind-phase luminosity is lower than the peak luminosity reported by the hydrodynamical model ($[1-2] \E{5} \Lsun$), and close to the Eddington limit ($\sim 9 \E{4} \Lsun$), since part of the total  energy outflow is now in the form of kinetic energy of the wind.\footnote{
  The stellar wind expands and cools the envelope, increasing opacity, which absorbs more radiative energy and decreases the photospheric luminosity.
  }  
  The luminosity drop at peak radial expansion could flatten the luminosity peak or create a double peak if significant.
  This flattening of the luminosity peak could indicate a stellar wind presence observationally.

  It is worth mentioning the similarity between the photospheric density and temperature time evolution, as well as their apparent inverse relation versus the photospheric radius. 
  In \citetalias{HSJ2023} we showed similar correlations related to a photospheric luminosity close to the Eddington limit. Correlations between observables such as photospheric luminosity and wind velocity, with the wind parameters determined by physical conditions of the inner parts of the envelope are shown in Table \ref{tab: Regression results}. The derivation of these correlations was first treated in \citetalias{HSJ2020}.
  
\begin{table}[htp]
  \begin{center}
  \caption{Regression results from correlations among observable variables and wind parameters in XRB-2.}
  \label{tab: Regression results}
  \scriptsize
  \begin{tabular}{c|ccc}
    Regression model   &   a   &   c   & $1-|\text{Corr}|$ \\
    \hline
    \rule{0pt}{2ex}
    $ \log T_\text{ph}    = \text{c} + \text{a} \log r_\text{ph} $
	&   $-0.504$   &   $10.45$   & $1\E{-6}$ \\
    $ \log \rho_\text{ph} = \text{c} + \text{a} \log r_\text{ph} $ 
	&   $1.014$    &   $1.233$   & $5\E{-6}$ \\
    $ \frac{8}{3} \frac{v_\text{ph}}{c} = \text{c} + \text{a} \frac{\dot M}{L_{R,\text{ph}}}\frac{G M}{ r_\text{ph}} $   
	&   $1.004$    &   $3\E{-5}$   & $3\E{-5}$ \\
    $  \frac{\dot E-L_{R,\text{ph}}}{L_{R,\text{ph}}} = \text{c} + \text{a} \frac{\dot M}{L_{R,\text{ph}}}\frac{G M}{ r_\text{ph}} $  
	&   $1.015$    &   $-1\E{-4}$   & $2\E{-4}$ 
  \end{tabular}
  \end{center}
\end{table}

\subsection{Observational signatures for NICER}
\label{sect: NICER simulation results}

With the aim of providing a first comparison with observational results, we constructed self-consistent time evolution curves of the photospheric parameters for the whole evolution of the four X-ray bursts studied in the present work, including both the wind-free phase (from the hydrodynamic model) and the radiative wind phase, as described in the last paragraph of Sect. \ref{sect: XRB_Wind}.
The resulting time evolution of effective temperature and bolometric luminosity were binned to 1-second bins, and used as input parameters for sequences of black-body spectral models in XSPEC. 
The black-body model was modified with TBabs to account for interstellar absorption. To provide simulations for a particular case, we take the interstellar absorption and distance corresponding to 4U 1820-30 ($N_H=1.3\times10^{22}$cm$^{-2}$ and distance of 6~kpc). 

\begin{figure*}
    \centering
    \def\figwidth{0.35\linewidth}
    \includegraphics[keepaspectratio,clip,trim=20 38 20 0,width=\figwidth]{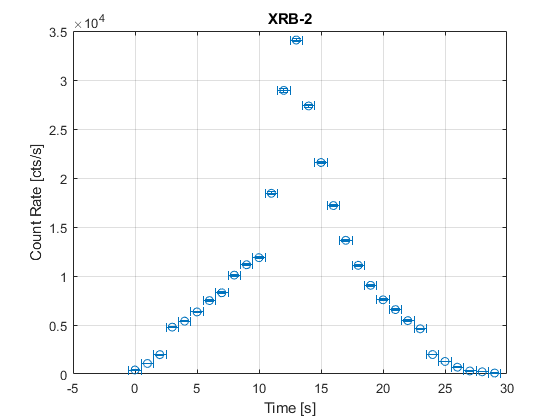}
    \includegraphics[keepaspectratio,clip,trim=20 38 20 0,width=\figwidth]{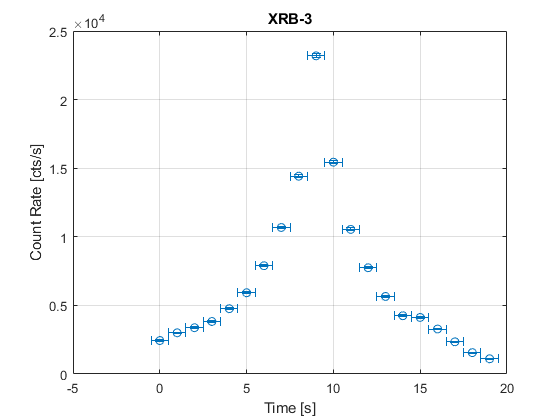}
    \\
    \includegraphics[keepaspectratio,clip,trim=20 0 20 28,width=\figwidth]{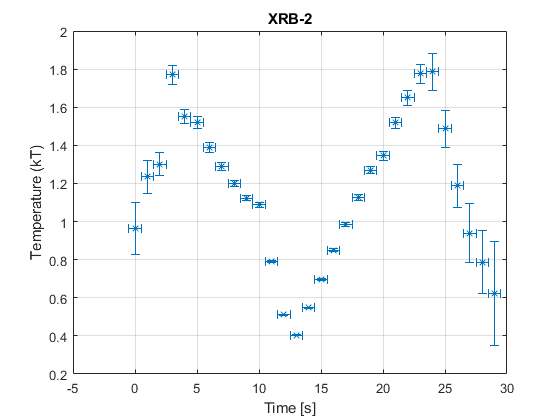}
    \includegraphics[keepaspectratio,clip,trim=20 0 20 28,width=\figwidth]{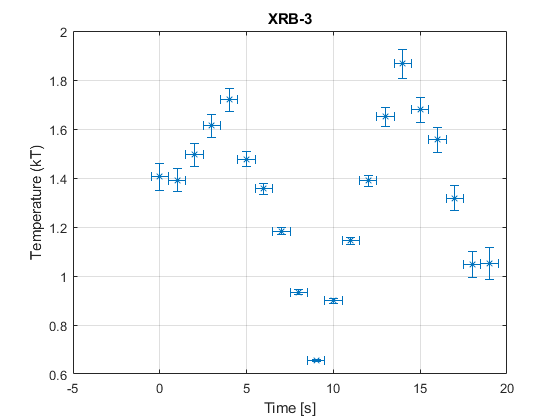}
    \caption{Photon count rate (top row panels) and effective temperature evolution (bottom row panels), simulated with \textit{XSPEC}, for XRB-2 and XRB-3 wind model predictions around peak expansion. The time starts at each burst onset.}
    \label{fig: XSPEC results}
\end{figure*}

Figure \ref{fig: XSPEC results} shows the simulated light-curve and effective temperature evolution during bursts 2 and 3. We must remark that, in our simulation, only the distance and the absorption data were made to correspond to 4U 1820-30, while the hydrodynamical simulations of the burst itself \citepalias{Jose2010}, which are the base of the results shown here, were not fine-tuned for this particular source. Still, even without using an XRB model that matches exactly the particular physical parameters of this source, we observe that the range of the simulated temperature variation (drop of a factor 3--4 at maximum expansion) and count-rate (peaking around 35000 cps) are compatible with the values observed for 4U 1820-30 (see for example Figs. 5 and 7 in \citealt{2024A&A...683A..93Y}, and Fig. 6 and 5 in \citealt{2024ApJ...975...67J}). Although the duration of the observed bursts for 4U 1820-30 is shorter than in our simulation, we can find other examples of double-peaked, expansion bursts with similar time-scales to our results (for example, for J1808.4–3658, figure 3 in \citealt{2019ApJ...885L...1B}.)

\section{Conclusions}\label{sect: conclusions}

 Our technique to link of the stellar wind model \citepalias{HSJ2020} to the XRB hydrodynamic models \citepalias{Jose2010} has allowed us to construct the time evolution of wind profiles and to quantify the mass-loss of each isotope produced by nucleosynthesis during the bursts. 
 In contrast to the single burst case we presented in \citetalias{HSJ2023}, where the technique was first introduced, in the present work we have shown the results for two consecutive bursts from  a sequence of four bursts in the XRB simulations (60-shell model in \citetalias{Jose2010}) which, albeit with a lower resolution, feature the whole evolution of the wind phase, as well as variations between consecutive bursts.

  The average ejected mass per unit time represents $0.3\%$ of the accretion rate, with $0.1\%$ of the envelope mass ejected per burst and $66\%-76\%$ of the ejecta composed by $^{60}$Ni, $^{64}$Zn, $^{68}$Ge, $^{58}$Ni, and $^{4}$He. 
  These are much lower values than the ones found in \citetalias{HSJ2023}. 
  Additionally, a noticeable variation of ejecta composition was observed between consecutive bursts. These differences are due to both the chemical evolution of the envelope and the depth of envelope layers blown away by the wind.
  Photospheric quantities show the same correlations found in \citetalias{HSJ2023}.

  Finally, we have built the complete evolution sequences (both wind and wind-free phases) of luminosity and effective temperatures for PRE bursts with radiative wind and used them to simulate \textit{NICER} observations. 
  We have found light-curve time-scales and dynamical ranges of count rates and temperatures compatible with actual NICER observations, for instance of 4U 1820-40 and J1808.4–3658.

  The present work establishes the bases for the future development of a grid of XRB-wind models, with varying NS and accretion parameters, that can be used as a diagnostic tool for PRE X-ray bursts showing evidence of radiative winds.
 This grid of models could help us fit the observed and predicted light-curves and temperature evolution curves to determine physical parameters of the wind, the X-ray bursts and the NS hosting them.
  

\section*{Acknowledgments}

 YH work at the Institute of Space Sciences is supported by Unidad de Excelencia Mar\'{i}a de Maeztu, grant CEX2020-001058-M, and Generalitat de Catalunya, grant 2021-SGR-1526, Spain. DMV, GS and JJ thank the Spanish MINECO grant PID2020-117252GB-I00 and the AGAUR/Generalitat de Catalunya grant SGR-386/2021. YC acknowledges support from the grant RYC2021-032718-I, financed by MCIN/AEI/10.13039/501100011033 and the European Union NextGenerationEU/PRTR.

\bibliography{biblio}%

\end{document}